\documentstyle[aps,pre]{revtex}
\begin{document}
\title{Third Cumulant of the total
Transmission of diffuse Waves}
\author{M.\ C.\ W.\ van Rossum, Johannes F.\ de Boer, and Th.\ M.\
Nieuwenhuizen}
\address{Van der Waals-Zeeman Laboratorium, Universiteit van Amsterdam
\\ Valckenierstraat 65, 1018 XE Amsterdam, The Netherlands}
\maketitle

\begin{abstract}
The probability distribution of the total transmission is studied for
waves multiple
scattered from a random, static configuration of scatterers.
A theoretical study of the second and third cumulant of this
distribution is presented. Within a diagrammatic approach a
theory is developed which relates
the third cumulant normalized to the average, $\langle \langle T_a^3 \rangle
\rangle$, to the normalized second cumulant $\langle \langle T_a^2 \rangle
\rangle$. For a broad Gaussian beam profile it is found that $\langle \langle
T_a^3 \rangle \rangle= \frac{16}{5} \langle \langle T_a^2 \rangle \rangle^2
$. This is in good agreement with data of optical experiments.
\end{abstract}

\today \\

\section{Introduction}\label{int}
Multiple scattering in disordered systems is a field of wide interest; it is
studied in electronic, microwave and optical systems. In the multiple
scattering regime the main transport in transmission is through diffusion.
Yet interference processes, possible by the underlying wave character, play
an important role. This interference leads to interesting effects such as the
enhanced backscatter cone\cite{kuga,albada3,wolf}, short-, and long-range
correlations\cite{stephen1,shapiro,feng,garcia,deboer}, and strong
localization.
If interference occurs between diffusion paths, it causes large fluctuations.
Most famous are the sample-to-sample fluctuations in the conductance of
electronic systems, the so called universal conductance
fluctuations\cite{altshuler2,lee,lee2}, but also other transmission
quantities are influenced by interference.

Recently, the attention is not only drawn to the variance of the
fluctuations, but to the entire distribution functions. Examples are the
intensity distribution in speckle patterns for classical waves
\cite{genack2,kogan} and the conductance distribution for electronic systems
\cite{altshulerboek}. The size of the fluctuations and the shape of the
distribution is related to the `distance' from the localization transition.
Far from localization, diffusion channels are almost uncorrelated and
fluctuations are small (except for the optical speckle pattern in the angular
resolved transmission). The correlation between the channels increases if the
localization is approached. The relevant parameter is the inverse
dimensionless conductance $1/g$, which can be interpreted as the chance that
two channels interfere. The dimensionless conductance can be expressed in the
thickness of the sample $L$, the mean free path $\ell$, and the number of
channels $N$, \begin{equation} g=\frac{4 N \ell}{3L}\label{eqdefg} \;
.\end{equation} The number of channels is calculated in analogy with a
waveguide, where it is unambigiously defined. In the diffuse mesoscopic
regime $g^{-1}$ will be a small parameter of our perturbation theory
(experimentally this proved fully justified, as there $g\sim 10^3$,
\cite{t3prl}). Close to Anderson localization $g$ approaches unity, and
fluctuations increase. The central question is how the distributions change
as the strong localization regime is approached\cite{altshulerboek,probprl}.

Let us briefly review some characteristics of optical transmission
distributions functions in the regime of moderate $g$. In the study of
mesoscopic systems using light scatterering, one takes a small sample with
static scatterers. In order to average, one needs to sum over a large number
of scatterer configurations, in practice this is done by varying an external
parameter such that the interference pattern is completely changed. Whereas
in electronic systems it is common to vary the Fermi energy or apply a
magnetic field, in optics one usually varies the wavelength of the light. In
contrast to electronic systems, not only the conductance but three different
transmission quantities can be measured in optical systems. (An exception is
the very recent observation of electron speckle by Gao {\it et al}.
\cite{gao}.) First, the angular resolved transmission can be considered: If a
laser shines on the sample, a speckle pattern is seen in the transmission. If
one measures the intensity in an outgoing direction $b$, this corresponds to
measuring the angular transmission coefficient $T_{ab}$, where $a$ denotes
the incoming and $b$ the outgoing channel. The intensities in the speckle
pattern have in zero'th order in $g$ an exponential distribution
\begin{equation} P(T_{ab})=\frac{1}{\langle T_{ab} \rangle
}\exp\left(-\frac{T_{ab} }{\langle T_{ab} \rangle} \right),
\label{eqray}\end{equation} which is also known as the Rayleigh law.
Deviations from this law occur if interference between the transmitted
intensities, the diffusons, is taken into account: the higher moments of the
distribution function increase. For large values of the transmission a
stretched exponential was observed\cite{genack2}, which was recently
predicted as\cite{probprl} $P(T_{ab})\propto \exp(-2\sqrt{g T_{ab}/\langle
T_{ab} \rangle }\;)$.

One obtains another transmission quantity, if one collects all the outgoing
light, and also all incoming directions are used, i.e. using a diffuse
incoming waves instead of a plane wave. In this case one summes over all
incoming and outgoing channels and the conductance of the
sample, $g= \sum_{a,b} T_{ab} $ is measured. It corresponds to the well
known conductance
measurement for electron systems. The conductance has roughly a Gaussian
distribution. The absolute variance is larger than expected classically
because of interference and it does not depend on the sample parameters
(hence the name universal conductance
fluctuations)\cite{altshuler2,lee,lee2}. Equivalently, the relative variance
is proportional to $g^{-2}$. The full distribution function of the
conductance was studied by Altshuler {\it et al.} \cite{altshulerboek}, who
predict a log-normal tail for the tail of the distribution function. For the
lower cumulants they predict that $\langle g^n \rangle_{\rm cum} \propto
\langle g ^{2-n} \rangle $. Note, however, that the prefactors maybe zero,
as Mac\^edo \cite{macedo} found that $\langle g^3 \rangle_{\rm cum} \propto
\langle g \rangle^{-2}$.

We are, however, interested in yet another transmission quantity. It is
obtained by spatially integrating the speckle pattern on the outgoing side,
and taking an incoming plane wave. One thus obtains an intermediate quantity
between the angular transmission $T_{ab}$ and the conductance $\sum_{ab}
T_{ab}$. It is termed the total transmission $T_a=\sum_b T_{ab} $. This
quantity is the subject of this paper. The total transmission is a constant
superposed with fluctuations. In first order of $g^{-1}$ the fluctuations
have a Gaussian distribution \cite{deboer,kogan}. The relative variance of
this distribution is proportional to $g^{-1}$, it is thus a factor $g$
larger than for the conductance fluctuations. This sensitivity of the total
transmission to interference processes and its simple limiting behavior (as
compared to the angular resolved transmission) make it an ideal quantity to
study mesoscopic transmission. Its full distribution was studied numerically
in 2D by Edrei {it et al.} who recovered the Gaussian distribution function,
tending towards a log-normal distribution near the Anderson
transition\cite{edrei}. The full distribution function was recently
derived\cite{probprl}. It was shown that it has a log-normal
distribution growth, and an exponential tail.

Recently the third cumulant of the distribution was
found experimentally by De Boer {\it et al.} \cite{t3prl}. In this paper we
present the theoretical
details of that work. We focus on the Gaussian distribution and the deviation
from the Gaussian due to the presence of the third cumulant.
The structure of this paper is as follows. The diffusion in optical systems
is described in section \ref{secdif}; in section \ref{seccum} the character
of the probability distribution is discussed. In section \ref{sect2} and
\ref{sect3}+\ref{secinf}, respectively, we calculate the second and third
cumulant of the probability distribution. Next we calculate experimental
corrections to our result in section \ref{secz0} and \ref{secjoh}, after
which we compare our results with the experimental data in section
\ref{secres}. We close with a discussion.

\section{Diffuse transport}\label{secdif}
Consider the transport of light  through a three dimensional slab with
static scatterers in  a random configuration. The slab has a thickness $L$.
Diffuse mesoscopic regime is characterized by $\lambda \ll \ell \ll L$. Here
$\ell$ denotes the mean free path and $\lambda$ denotes the wavelength in
the medium. Isotropic scattering is assumed. We also work in the scalar wave
approximation. It is known that the two independent polarization directions
of light effectively double the number of channels $N$ in the problem as
compared to the scalar wave case\cite{deboer}. Apart from a small
correction term, our final result does not depend on the number of channels
and the doubling.

We take the depth in the the slab as the $z$-coordinate, the slab thus
corresponds to $0<z<L$. A plane scalar wave with unit flux and area $A$
impinges on the sample from direction $a$. It is given by
\begin{equation} \psi^a_{\rm in}({\bf r})=\frac{1}{\sqrt{Ak
\mu_a}}\exp(i{\scriptstyle P}_a{\scriptstyle R}+ik
\mu_a z), \qquad z<0, \label{eqplanewave}\end{equation} where $k$ is the
wavenumber, ${\scriptstyle R}=(x,y)$ is the transversal coordinate,
${\scriptstyle P}_a$ is the transverse component of the momentum, and
$\mu_a=\sqrt{1-{\scriptstyle P}_a^2/k^2}=\cos\theta_a$, where $\theta_a$ is
the angle with respect to the $z-$axis. The number of channels depends on the
area of the incoming beam. Dividing the total area of the beam with diameter
$\rho_0$ in small coherence regions of area $\lambda^2$, one obtains in
analogy with waveguides $N={\it 2} \times k^2\rho_0^2 /4$, or the Weyl
formula $N={\it 2} \times k^2 A/4\pi$. The factor ${\it 2}$ comes from the
two independent polarization directions. For the
case of scalar wave scattering this doubling factor should be left out.
The main contribution to the average transmission is given by the
diffuse transport of intensity. This means that the two amplitudes, which
make up the intensity, scatter on the same scatterers following the same path
through the sample. In a diagrammatic language these processes are known as
diffusons or ladder diagrams. An example of such a scattering process in
drawn in Fig.\ref{figdif}, in the following we will depict the diffuson by
close parallel lines and omit the scatterers for clarity.

In the bulk the diffuse intensity obeys simply the diffusion equation, but
the precise prefactors from the coupling to the outside
have to be determined from the Schwarzschild-Milne equation
\cite{vanderhulst2}
\begin{equation} {\cal L} (z)=S(z)+\int_0^L {\rm d}z' M(z,z') {\cal L}(z').
\label{eqmilne}\end{equation} This is a
self-consistent transport equation which gives the diffuse intensity ${\cal
L}$ resulting from a source $S$. The kernel $M$ describes the intensity decay
between two scatterings. In the ladder approximation (without internal
reflections) it is \begin{equation} M(z,z')=\int_0^1
\frac{{\rm d}\mu}{2\mu} {\rm e}^{-|z-z'|/\mu \ell}. \end{equation} The source
term $S$ is given by the single scattered incoming intensity.
Our incoming plane
wave leads to a source  in the Milne equation \begin{equation} S(z)=
 nt\overline{t}|\Psi_{\rm in}^a|^2= \frac{4\pi}{A k \mu_a
\ell}
\exp(-z/(\mu_a \ell)) .\end{equation} where $n$ is scatterer density, $t$ is
the $t-$matrix, giving the mean free path $\ell=4\pi /(nt \overline{t})$.
The transport equation (\ref{eqmilne}) now
yields for an incoming
diffuson \begin{equation} {\cal L}^a_{\rm in}( z)=
 \frac{4\pi
\tau_1(\mu_a)} {k\ell A\mu_a}\,\frac{L-z}{L}, \label{eqdifin}
\end{equation} where $\tau_1$
describes the limit intensity of a semi-infinite system, see Ref.~\cite{thmn5}.

Now the prefactors are known, we continue with the diffusion equation,
which holds a few mean free paths away from the
surface.
We generalize to the case where the amplitudes making up the diffuson have a
non-zero momentum. For the corresponding diffusion
propagator propagating from the point $z$ to the point $z'$ it holds that
 \begin{equation} -\nabla^2 {\cal
L}^{\rm int}(z,z') = -\partial_z^2 {\cal L}^{\rm int}(z,z')+{\scriptstyle
Q}^2  {\cal L}^{\rm int}(z,z')
=\frac{12 \pi}{\ell^3} \delta(z-z'), \label{eqdifeq} \end{equation} with
approximate boundary conditions ${\cal L}^{\rm int}(0,z';{\scriptstyle
Q})={\cal L}^{\rm int}(L,z';{\scriptstyle Q})=0$. Here ${\scriptstyle Q}$ is
a two-dimensional vector, defined  as the difference of perpendicular
momenta of the two incoming amplitudes
making up the diffuson. If coherent waves impinge with a different angle on
the sample, this net transverse momentum of diffusons need not be zero and
the diffusons decay exponentially with the inverse decay length  equals
$|{\scriptstyle Q}|$. The solution of the diffusion equation reads
\begin{equation} {\cal L}^{\rm int}(z,z';{\scriptstyle Q})=
\frac{12\pi}{\ell^3} \frac{\sinh(|{\scriptstyle Q}| z_<) \sinh(|{\scriptstyle
Q}|(L-z_>))}{|{\scriptstyle Q}| \sinh(|{\scriptstyle Q}| L)}, \label{eqladQ}
\end{equation} with $ z_< = \min(z,z')$ and $z_>= \max(z,z')$.
Below we also need incoming diffusons with transverse momentum, they are
obtained by combing the prefactor of Eq.~(\ref{eqdifin}) with the $z_>$
dependence of Eq.~(\ref{eqladQ}).

The
total transmission is obtained by integrating over the outgoing channels.
In
the experimental geometry of Ref.~\cite{t3prl}, the outgoing radiation was
collected in a
integrating sphere. Only outgoing diffusons where its amplitudes are exactly
in phase (i.e. have opposite phase) are leading after this integration.
Therefore
the outgoing diffuson can have no transverse momentum. The transport equation
(\ref{eqmilne}) yields an outgoing diffuson from a unit source
\begin{equation} {\cal L}_{out}=\frac{k}{\ell }\frac{z}{L}.
\label{eqdifout} \end{equation}
The total transmission is obtained by connecting the incoming to the outgoing
diffuson, as was shown in Refs.~\cite{thmn5,c3prb} \begin{equation} \langle
T_a \rangle = {\it 2}
\times \frac{\tau_1(\mu_a) \ell}{3 L \mu_a} \label{eqtaav}.\end{equation}
Here we included the doubling from the two polarization directions.
Also
integrating over all incoming directions yields the conductance\cite{c3prb}
\begin{equation} g=\sum_a \langle T_a\rangle= {\it 2} \times \frac{\ell k^2
A}{3\pi L},\end{equation} with $N={\it 2} k^2 A/4\pi $ this equals our
previous definition (\ref{eqdefg}).

\section{Cumulants of the probability distribution} \label{seccum}

In this section we introduce the probability distribution of the total
transmission of scalar waves and we discuss some of its properties. The
corrections for vector waves will be made in section \ref{secjoh}. We will
link the moments
of the distribution to diagrams. The moments of the probability distribution
$P(T_a)$ can be extracted as \begin{equation} \langle T_a^k \rangle =\int
{\rm d}T_a \, P(T_a)
T_a^k. \end{equation} In a diagrammatic approach the $k$-th moment can be
represented by
a diagram with $k$ diffusons on both incoming and outgoing side. The $k=1$
term is the average total transmission $\langle T_a \rangle$, as given by
the Schwarzschild-Milne equation in
Eq.(\ref{eqtaav}). This quantity is given by a single diffuson and is thus
independent of channel-to-channel correlations. The second moment can be
decomposed in the first two cumulants: \begin{equation} \frac{ \langle
 T_a^2 \rangle }{\langle T_a \rangle^2} =\frac{\langle T_a\rangle^2 +\langle
T_a^2\rangle_{\rm cum} }{\langle T_a \rangle^2}=1 +\langle \langle T_a^2
\rangle
\rangle , \end{equation} the double brackets denote cumulants normalized to
the average.
Diagrammatically the second moment is depicted in Fig.\ \ref{figt2}.  The
decomposition in cumulants will prove useful as each cumulant corresponds to
a different number of interactions between the diffusons.  In the first
term, Fig.\ \ref{figt2}(a) there is no interference; it factorizes in the
average transmission squared (apart from a small correction discussed in
section \ref{secjoh}). The second term, Fig.\ \ref{figt2}(b), is the second
cumulant $\langle \langle T_a^2 \rangle \rangle$. It gives the variance of
the fluctuations. Interactions between two diffusons are responsible for the
presence of this  second cumulant.  The interference process is in a
diagrammatical language the so-called Hikami-box, it is depicted as the
shaded square in the figure. In the box two amplitudes of the incoming
diffusons are interchanged, causing a correlation between the outgoing ones.
Precisely the same process was studied in the more general context of the
long range correlation functions\cite{pnini2,spivak,berkovits3,albada1}. In
the work of De Boer {\it et al.} \cite{deboer} and Garcia {\it et al.}
\cite{garcia} the correlation between two
diffusons with different frequency was measured. This is the so called $C_2$
correlation function: $ C_2(\Delta \omega)=\langle \langle T_a(\omega)
T_a(\omega+\Delta \omega) \rangle \rangle$. For our case find that $ \langle
\langle T_a^2 \rangle \rangle=C_2(0) $ and thus corresponds to the peak
value of
this correlation function. In the next section we will calculate this
cumulant in detail.

Similar to the second moment, one can distinguish three different
contributions to the third moment, \begin{equation} \frac{ \langle
T_a^3 \rangle}{\langle T_a \rangle^3} =1+3 \langle \langle T_a^2 \rangle
\rangle + \langle \langle T_a^3 \rangle \rangle. \label{t3=} \end{equation}
The corresponding leading diagrams are drawn in Fig.\ \ref{figt3}. The first
term, Fig.\ \ref{figt3}(a), again corresponds to the transmission without
interference. The second term, Fig.\ \ref{figt3}(b), is reducible in a single
diffuson and a second cumulant diagram. From the figure it is clear that this
decomposition can be done in three ways which is reflected in the prefactor
of $\langle \langle T_a^2 \rangle \rangle $ in Eq.~(\ref{t3=}). The third
contribution stands for the third cumulant in the distribution and expresses
the leading deviation from the Gaussian distribution. This is the term we are
mainly interested in. It consists of two related diagrams: Fig.\
\ref{figt3}(c+d). The three intensities can interfere twice two by two, or
the intensities can interact all three together, with a so called Hikami
six-point vertex. Both contributions will prove to be of the same order
of magnitude. The strength of the effect can be easily estimated using the
interpretation of $1/g$ as a interaction probability. By looking at the
diagram, the third cumulant is proportional to the chance of two diffusons
meeting twice, thus of the order $1/g^2$. We find the basic result
\begin{equation} \langle \langle T_a^3 \rangle \rangle \propto \langle
\langle T_a^2 \rangle \rangle^2.  \end{equation} The rest of the paper
essentially consist of proving this relation and determination of the
prefactor. Finally we will compare this relation to the experimental data of
Ref.~\cite{t3prl}.

Because only a limited number of channels is sampled in an experiment, the
law of large numbers predicts a distribution with some non-zero width even if
we only would consider disconnected diagrams. However, we will show in
section \ref{secjoh} that this effects brings only a negligible contribution
to the measured cumulants. The large fluctuations are almost only due to the
interference.

Note that in our calculations diagrams with loops are neglected.  An example
of a loop diagram is the $C_3$ (or universal conductance fluctuation)
contribution to the second cumulant\cite{lee,feng,c3prb}. One of the $C_3$
diagrams is a second cumulant diagram where the outgoing diffusons are again
input for another Hikami-box. (It is like glueing two second cumulant
diagrams after each other.) This diagram contains two Hikami boxes, therefore
it gives a contribution of order $g^{-2}$ to the second cumulant. In general,
one easily sees that in order to create a loop, one needs a higher number of
interference vertices. Therefore, these diagrams are of higher order in
$1/g$, and we did not calculate them. The leading contributions to the
cumulants are by far sufficient for the description of the experiment of
Ref.~\cite{t3prl}.

\section{The second cumulant}\label{sect2}
In this section we recover the results for the
second cumulant. This quantity was often calculated in literature.
In the work of the Boer {\it et al.}\cite{deboer} the frequency correlation
in the total transmission was calculated using a Langevin approach
introduced by Spivak and Zyuzin\cite{spivak}, see also Pnini and
Shapiro\cite{pnini2}. Here we obtain the same results using a diagrammatic
technique\cite{feng,stephen1,skin}. We calculate the diagram of
Fig.\ref{figt2}(b).
The interaction vertex of the diffusons is the Hikami four point
vertex or Hikami box\cite{hikami}.
The Hikami box is depicted in Fig.\ref{figh4sob}; from the figure it is
easily seen that the vertex interchanges two amplitudes of the incoming
diffusons. As is shown in Fig.\ref{figh4sob}, the full vertex is (in second
order Born approximation) the sum of one bare vertex and two vertices with an
additional scatterer. (If  the scatterers are near resonance, the second
order Born approximation is no longer  valid and more diagrams are relevant.
However, for the four point vertex it was explicitly proven in
Ref.\cite{hik} that this will only be reflected in the different value of
the mean free path.) The summation of the diagrams yields\cite{hikami,gorkov}
\begin{equation} H_4 =\frac{-\ell^5}{48 \pi k^2}\left[{\bf q}_a\! \cdot \!
{\bf q}_{a'}+{\bf q}_b \! \cdot \! {\bf q}_{b'}-\frac{1}{2}({\bf q}_a^2+{\bf
q}_{a'}^2+{\bf q}_b^2+{\bf q}_{b'}^2) \right]  \label {haha} .\end{equation} To
this vertex two incoming diffusons, $a$ and $a'$, and two outgoing diffusons,
$b$ and $b'$, are attached. The ${\bf q}_i$ denotes the {\em three
dimensional} momentum of the corresponding diffuson. The momenta are pointing
towards the vertex. For our second cumulant diagram of Fig.\ref{figt2}(b) we
have numbered $a=1$, $a'=3$, $b=2$ and $b'=4$.

As stated above, only the outgoing diffusons where the amplitudes have
exactly opposite phase are leading in the total transmission measurements.
Therefore the transversal momenta of the outgoing diffusons must be zero
${\scriptstyle Q}_2={\scriptstyle Q}_4=0$, but as amplitudes are exchanged in
the Hikami-box, the incoming diffusons can have a transversal component in
their momentum. Using momentum conservation one has ${\scriptstyle
Q}_1=-{\scriptstyle Q}_3\equiv {\scriptstyle Q}$

After Fourier transforming
Eq.~(\ref{haha}) in the $z-$coordinate we find
\begin{equation} H_4
=\frac{\ell^5}{48 \pi k^2}\left(\partial_{z_1} \partial_{z_3}
+\partial_{z_2}
\partial_{z_4}-{\scriptstyle Q}_1\! \cdot \!  {\scriptstyle Q}_3
-\frac{1}{2} \sum_{i=1}^4
(\partial_{z_i}^2-{\scriptstyle Q}_i^2) \right)  \label {haha2} .\end{equation}
Here $\partial_{z_1}$ stands for differentiating towards the $z-$coordinate
of the diffuson with the same number, in this case  ${\cal L}_1(z)$.
The
terms inside the sum in Eq.(\ref{haha2}) are neglected. According to
the diffuson equation, they yield approximately a source term near the
boundary of the sample. In the integral these contributions are of the
negligible order.
In the plane wave limit of the incoming beam,
all transverse momenta are absent and the diffusons are simple linear
functions, given by Eq.~(\ref{eqdifin}) and (\ref{eqdifout})..
One obtains the known result for the second cumulant\cite{stephen1}
\begin{eqnarray} \langle \langle T_a^2 \rangle \rangle&=& \langle T_a
\rangle^{-2} \int \int {\rm d}x {\rm d}y  \int_0^L {\rm d}z H_4 {\cal L}_1(z)
{\cal L}_2(z) {\cal L}_3(z) {\cal L}_4(z)\nonumber \\&=&\frac{1}{gL^3}
\int_0^L {\rm d}z [z^2+ (1-z)^2]= \frac{2}{3} g^{-1} .\end{eqnarray}
Taking only this second cumulant we find for the moment a
Gaussian probability distribution in the plane wave limit
\cite{kogan}
\begin{equation} P(T_a) = \sqrt{\frac{3g}{4\pi}} \exp\left[ -\frac{3g}{4}
\left(T_a-\langle T_a \rangle\right)^2 \right] \label{eqpta}.\end{equation}

Let us now study the influence of the beam profile on the correlation. If
the spot of the incoming beam is finite, amplitudes
with different transverse momenta are present. They can combine into
incoming diffusons with perpendicular momentum. We suppose that the
incoming beam has
a Gaussian profile. It is decomposed into plane waves defined in
Eq.~(\ref{eqplanewave}) (for convenience we assume perpendicular incidence)
\begin{equation} \psi_{\rm in}=\frac{2\pi}{W} \sum_a \phi({\scriptstyle
Q}_a) \psi_{\rm in}^a ,\qquad  \phi({\scriptstyle Q})
=\frac{\rho_0}{\sqrt{2\pi} }{\rm e}^{-{\scriptstyle Q}^2 \rho_0^2/4}
\label{eqpsig},\end{equation}
where $\rho_0$ is the beam diameter. In order to have two diffusons with a
momentum ${\scriptstyle Q}$ and $-{\scriptstyle Q}$,
we find that the four incoming amplitudes combine to a weight function
\begin{equation} \int {\rm d}^2{\scriptstyle P}_1 {\rm d}^2{\scriptstyle P}_3
\;
\phi({\scriptstyle P}_{1}) \phi^\ast({\scriptstyle P}_1+{\scriptstyle Q})
\; \phi({\scriptstyle P}_{3}) \phi^\ast({\scriptstyle P}_3-{\scriptstyle Q})
={\rm e}^{-\rho_0^2 {\scriptstyle Q}^2 /4}. \end{equation}
The second cumulant is now found by first calculating the ${\scriptstyle
Q}-$dependent correlation. This is done similar to the calculation above, but
now the ${\scriptstyle Q}_1\! \cdot \! {\scriptstyle Q}_3$-terms in
Eq.(\ref{haha2}) should be taken into account, as well as the ${\scriptstyle
Q}$-dependence of the incoming diffusons, see Eq.(\ref{eqladQ}). (The
outgoing diffusons still have no such momentum.) The total cumulant is now
found by integrating over the momentum with the corresponding weight. For a
Gaussian beam profile one finds\cite{deboer} \begin{equation} \langle \langle
T_a^2 \rangle \rangle= \frac{\rho_0^2}{4\pi g} \int {\rm d}^2{\scriptstyle Q}
\; {\rm e}^{-\rho_0^2 {\scriptstyle Q}^2 /4} F_2(|{\scriptstyle Q}| L)
\label{c2} ,\end{equation} with $F_2(x)=[\sinh(2x)-2x]/[2x \sinh^2 x]$. If the
incoming beam is again very broad, $\rho_0 \gg L$, only the term
$F_2(|Q|L=0)=2/3$ contributes and one recovers the plane wave behavior
$\langle \langle T_a^2 \rangle \rangle= 2/3g $. Note that this agreement is
found by identifying  the area of a Gaussian profile with $A=\pi\rho_0^2$.
This definition is somewhat arbitrary and other choices are
also possible\cite{probprl}. After fixing this definition, no freedom remains
and we will see that for the third cumulant a Gaussian
profile leads to other results than a plane wave.

The second cumulant decreases as $1/\rho_0^2$ at large $\rho_0$. In a real
space picture it is evident that the correlation increases if the two
incoming channels are closer to each other, i.e. if the beam diameter is
smaller. In the experiment of Ref.\cite{t3prl} the focus was kept small in
order to minimize the dimensionless conductance $g$ and therefore to maximize
the fluctuations.

\section{The third cumulant}\label{sect3}
We now discuss the calculation the third cumulant. As mentioned, there are
two processes contributing. One with two four point vertices which we term $
\langle \langle T_a^3 \rangle \rangle_c$ and one with a six point vertex $
\langle \langle T_a^3 \rangle \rangle_d$, where we have chosen the subscript
according to Fig.~\ref{figt3}. This process was, to our knowledge, not
studied before. Our calculation follows the lines of the second
cumulant calculation.

\subsection*{Interference via two four point vertices}

First consider the diagram in Fig.~\ref{figt3}(c). We have
labeled to incoming diffusons with odd numbers, the outgoing ones with
even numbers. Two incoming
diffusons, ${\cal L}_1$ and ${\cal L}_3$ meet at a position $z$. In a Hikami
box the diffusons ${\cal L}_1$ and ${\cal L}_3$ interfere into ${\cal L}_2$
and an internal diffuson ${\cal L}^{\rm int}_{78}$. The ${\cal L}_2$
propagates out, whereas ${\cal L}^{\rm int}_{78}$ interferes again at $z'$
with incoming diffuson ${\cal L}_5$ into two outgoing ones ${\cal L}_4$ and
${\cal L}_6$. Apart  from this process, also three other sequences of
interference are possible. This means that the diffusons can also be permuted
as: $({\cal L}_1,{\cal L}_3, {\cal L}_5,{\cal L}_2,{\cal L}_4,{\cal L}_6)
\rightarrow ({\cal L}_3,{\cal L}_5, {\cal L}_1,{\cal L}_4,{\cal L}_6,{\cal
L}_2) \rightarrow ({\cal L}_5,{\cal L}_1,{\cal L}_3,{\cal L}_6,{\cal
L}_2,{\cal L}_4)$. We will denote the sum over these permutations as
$\sum_{\rm per}$. As the diagrams can also be complex conjugated, there is
also an combinatorical factor 2 for all diagrams. (Note that complex
conjugation for the {\em second} cumulant diagram, yields no different
diagram and thus it should not be taken into account.)
The expression for the diagram of Fig.~\ref{figt3}(c) is now
\begin{equation} \langle \langle T_a^3\rangle
\rangle_{c}=\langle T_a\rangle^{-3} \, 2\sum_{\rm per} A \int_0^L {\rm d}z
\int_0^L
{\rm d}z' \; H_4(z) H_4(z') {\cal L}_1(z) {\cal L}_2(z) {\cal L}_3(z) {\cal
L}_4(z') {\cal L}_5(z') {\cal L}_6(z') {\cal L}_{78}^{\rm int}(z,z')
.\end{equation}
In turns out that it is useful to rewrite the form of the Hikami boxes as
introduced in Eq.(\ref{haha}) into an equivalent expression, using momentum
conservation ${\bf q}_a+{\bf q}_{a'}+{\bf q}_{b}+{\bf q}_{b'}=0$. In real
space the use of momentum conservation corresponds to partial integration.
 The Hikami box  is again simplified using the fact that there are no
transversal
 momentum terms, or ${\scriptstyle Q}$ terms, for the outgoing diffusons.
Also the source terms
$q^2$  of incoming and outgoing diffusons are neglected. Using the numbering
in Fig.~\ref{figt3}, we obtain
\begin{equation}
H_4(z)=\frac{- \ell^5}{48\pi k^2}[ 2\partial_{z_1}\partial_{z_2}+
2 \partial_{z_2}\partial_{z_3} ]
, \qquad H_4(z')=\frac{\ell^5}{48\pi k^2}[2\partial_{z_4}\partial_{z_6}-
\partial_{z_8}^2 + {\scriptstyle Q}_8^2]. \label{eqh4keus} \end{equation}
Source terms, i.e. $q_i^2-$terms of the incoming and outgoing diffusons were
again neglected, but the source term of the
diffuson between the vertices is important. As is seen with the diffusion
equation (\ref{eqdifeq}), it brings  \begin{equation} \partial_{z_8}^2 {\cal
L}^{\rm int}_{78} (z,z')
+{\scriptstyle Q}_8^2 {\cal L}^{\rm  int}_{78}(z,z') =\frac{\ell^3}{12\pi}
\delta(z-z').\end{equation} The contribution
from the
source term (i.e. $H_4(z')\propto -\partial_{z_8}^2 +
{\scriptstyle Q}_8^2$,
$H_4(z) \propto \partial_{z_1}\partial_{z_2}+\partial_{z_2}\partial_{z_3}$)
 is  \begin{eqnarray}
&& -\left(\frac{\ell^5}{48\pi k^2} \right) ^2 4\sum_{\rm per}
A\int {\rm d}z (\partial_{z_1}  \partial_{z_2}+\partial_{z_2}\partial_{z_3})
{\cal L}_1 {\cal L}_2 {\cal L}_3
\int {\rm d}z' (-\partial_{z_8}^2  + {\scriptstyle Q}_8^2) {\cal L}^{\rm
int} {\cal L}_4 {\cal L}_5 {\cal L}_6 \nonumber \\
&=& -\frac{\ell^7 A}{48\pi k^4} \int_0^L {\rm d}z [\partial_{z_1}
\partial_{z_2}+\partial_{z_2}\partial_{z_3} +\partial_{z_3}
\partial_{z_4}+\partial_{z_4} \partial_{z_5}+\partial_{z_5}
\partial_{z_6}+\partial_{z_6} \partial_{z_1}] \times \nonumber \\ &&{\cal
L}_1 {\cal L}_2 {\cal L}_3{\cal L}_4{\cal L}_5
{\cal L}_6 \label{h4bron}. \end{eqnarray} Although this corresponds to a
local process (just one $z-$coordinate is involved), it is of leading order.
Together with the expression coming from $H_4(z')$ proportional to
$\partial_{z_4}\partial_{z_6}$,
we find for the total contribution of the process in  Fig.\ref{figt3}(c)
\begin{eqnarray} \langle \langle  T_a^3 \rangle \rangle_{c} &=& \langle T_a
\rangle^{-3}
\left(\frac{l^5}{48 \pi k^2} \right)^2 8\sum_{\rm per}  A\int_0^L {\rm d}z\;
{\cal L}_1(z){\cal L}_2'(z){\cal  L}_3(z)\int_0^L {\rm d}z'\; {\cal
L}_4'(z'){\cal L}_5(z'){\cal L}_6'(z')
\partial_z{\cal L}^{\rm int} (z,z') \nonumber \\
&&  -\langle T_a \rangle^{-3} \frac{\ell^7 A}{48\pi k^4} \int_0^L {\rm d}z
[\partial_{z_1} \partial_{z_2}+\partial_{z_2}\partial_{z_3} +\partial_{z_3}
\partial_{z_4}+\partial_{z_4} \partial_{z_5}+\partial_{z_5}
\partial_{z_6}+\partial_{z_6} \partial_{z_1}] \times \nonumber \\ &&{\cal
L}_1(z) {\cal L}_2(z)
 {\cal L}_3(z) {\cal L}_4(z) {\cal L}_5(z) {\cal L}_6(z) \label{eqt3hh},
\end{eqnarray}
where ${\cal L}'(z)$ denotes the derivative towards $z$ of ${\cal L}(z)$.
Calculated for a plane wave it gives  \begin{equation} \langle \langle
T_a^3 \rangle
\rangle_{c}=\frac{22}{15g^2},\end{equation} which is indeed proportional to
$g^{-2}$, as predicted.

\subsection*{Contribution of six point vertex}
There is another diagram contributing to the third cumulant which is of the
same order of magnitude as the process calculated above; it is depicted in
Fig.~\ref{figt3}(d). It can be thought of the following way: The use of
Hikami box in the previous section assumes that the outgoing legs scatter at
least once before they propagate out or interfere again. This is a reasonable
assumption for the outgoing diffusons, but for the internal diffuson ${\cal
L}^{\rm int}_{78}$ it also possible that coming from $z$ it directly, i.e.
without scattering, interferes again at $z'$. This process is not included in
the calculation of the previous section but has to be studied separately. The
unscattered intensity decays exponentially over one mean free path, therefore
this process is only important if $z$ and $z'$ are within one mean free path.
We denote the corresponding vertex as $H_6$, as six diffusons are connected
to this diagram. This diagram was also already calculated by
Hikami\cite{hikami}. Also here, the dressings of the diagrams have to be
added to the bare diagrams. Taking rotations of the depicted diagrams into
account, there are 16 diagrams in second order Born approximation. It is not
allowed to dress the bare six-point vertex (leftmost r.h.s. diagram in Fig.\
\ref{figh6}) with a scatterer which connects two opposite propagators. This
dressing gives  also a leading contribution even if the dressing performed
with an
arbitrary number of scatterers, but the resulting diagram is the same as the
composed diagram  with two four point vertices, Fig.~\ref{figt3}(c), and
should thus not be counted.
Yet this observation is useful to check the combinatorical ratio between the
six point vertex and the composed diagram: the forbidden dressing can be
performed in three ways. As the diagrams can also be complex conjugated,
there is also a factor 2 for all diagrams.

In the lowest order of $(q\ell)$ we find for the six point vertex
\begin{equation} H_6=
\frac{-\ell^7}{96\pi k^4} [{\bf q}_1 \! \cdot \!  {\bf q}_2+{\bf q}_2 \!
\cdot \!  {\bf q}_3+{\bf q}_3 \! \cdot \!  {\bf q}_4+{\bf q}_4 \! \cdot \!
{\bf q}_5+{\bf q}_5 \! \cdot \!  {\bf q}_6+{\bf q}_6 \! \cdot \!  {\bf q}_1
+\sum_i q_i^2] .\end{equation} Hikami's original expression
can be recovered from this using momentum conservation.
After a Fourier-transformation in the $z-$direction the six-point vertex
yields a contribution to the third cumulant \begin{eqnarray}
\langle\langle{T_a^3} \rangle\rangle_{d}&=& \langle T_a \rangle^{-3}
\frac{\ell^7 A}{48\pi k^4} \int_0^L {\rm d}z [\partial_{z_1}
\partial_{z_2}+\partial_{z_2}\partial_{z_3} +\partial_{z_3}
\partial_{z_4}+\partial_{z_4} \partial_{z_5}+\partial_{z_5}
\partial_{z_6}+\partial_{z_6} \partial_{z_1}] \times \nonumber \\ &&{\cal
L}_{1}(z)
{\cal L}_{2}(z) {\cal L}_{3}(z) {\cal L}_{4}(z) {\cal L}_{5}(z)   {\cal
L}_6(z) .\end{eqnarray} Here we used the fact that all outgoing diffusons
have zero transversal momentum. Therefore all ${\scriptstyle Q}_i
{\scriptstyle Q}_j$-terms are absent. In the limit of an incoming plane wave
we find a contribution to the third cumulant \begin{equation} \langle \langle
T_a^3 \rangle \rangle_{d}=-\frac{2}{5g^2}. \end{equation}
The contribution from the source term, i.e. Eq.(\ref{h4bron}) of the
previous section, exactly cancels the
contribution from the six-point vertex. The cancellation
 seems plausible as one does not
expect short distances properties to be important in the total process.
Nevertheless,
this cancellation depends on the precise form of the Hikami four-point
vertex in Eq.(\ref{eqh4keus}). If we use other equivalent forms of the
Hikami-box
the contributions of the single and double integral in Eq.(\ref{eqt3hh}) are
shifted with respect to each other and a
full cancellation is not present. Of course, neither the result for
Eq.(\ref{eqt3hh}) nor the final result for $\langle \langle T_a^3 \rangle
\rangle$
relies on this choice. The precise mechanism behind this is not clear to us.
However, using the cancellation we
only need to consider the term in Eq.(\ref{eqt3hh}), which comes from
$H_4(z) \propto \partial_{z_1}\partial_{z_2}+\partial_{z_2}\partial_{z_3}$,
$H_4(z') \propto \partial_{z_4}\partial_{z_6}$ and the permutations.
We thus obtain for the third cumulant  \begin{eqnarray} \langle \langle
T_a^3 \rangle
\rangle&=& \langle \langle T_a^3 \rangle \rangle_{c} +\langle \langle T_a^3
\rangle \rangle_{d}\nonumber \\ &=& \langle T_a \rangle^{-3}
\frac{\ell^{10} A}{288 \pi^2 k^2} \sum_{\rm per} \int_0^L {\rm d}z \;
{\cal L}_1(z){\cal L}_2'(z){\cal L}_3(z)
\int_0^L {\rm d}z'\; {\cal L}_4'(z'){\cal  L}_5(z'){\cal L}_6'(z')
\partial_z{\cal L}^{\rm int}(z,z') \label{eqt3}
.\end{eqnarray} In the next section we calculate this expression for
various incoming beam profiles.

\section{Influence of incoming beam profile}
\label{secinf} Now that the leading interference processes are known, inserting
the diffusons gives the final value of the third cumulant. We first consider
the simple case of incoming plane waves. As there can be no transversal
momentum difference in the incoming amplitudes, all ${\scriptstyle Q}_i$
vanish.  As a result
all diffusons are simple linear functions of $z$. And we find from
Eq.~(\ref{eqt3})\begin{equation} \langle \langle T_a^3 \rangle
\rangle=\frac{16}{15g^2}=\frac{12}{5} \langle \langle T_a^2 \rangle \rangle^2
\qquad \mbox{plane wave;} \; \; \rho_0 \gg L.\label{eqt3pw} \end{equation}

In practice, however, we deal with a Gaussian beam with limited spot size,
influencing the cumulants in two ways. First, if the spot size decreases to
values comparable to the sample thickness we have to convolute over a range
of incoming momenta, just like we did when calculating the second cumulant.
Second, the Gaussian profile brings an extra geometrical factor as we
will show below.

We need the expression when diffusons with arbitrary momentum are connected
to the Hikami boxes. The outgoing diffusons still have no transverse momentum.
Because of momentum conservation, the transversal momentum ${\scriptstyle
Q}_7$ of the
diffuson connecting the two four-boxes must equal ${\scriptstyle Q}_5$.
The integration over the possible momenta results again
in a Gaussian weight function. From the definition (\ref{eqpsig}) we derive
\begin{equation}
\int {\rm d}^2{\scriptstyle P}_1 {\rm d}^2{\scriptstyle P}_3 {\rm
d}^2{\scriptstyle P}_5 \;
\phi({\scriptstyle P}_1) \phi^\ast({\scriptstyle P}_1+{\scriptstyle Q}_1) \;
\phi({\scriptstyle P}_3) \phi^\ast({\scriptstyle P}_3+{\scriptstyle Q}_3) \;
\phi({\scriptstyle P}_5) \phi^\ast({\scriptstyle P}_5+{\scriptstyle Q}_5)
={\rm e}^{-\rho_0^2({\scriptstyle Q}_1^2+
{\scriptstyle Q}_3^2+{\scriptstyle Q}_5^2)/8} \end{equation}
Momentum conservation is used to eliminate also ${\scriptstyle Q}_5$ and
reduce the integration to
two transversal momenta. The final result for the third cumulant is obtained
by
inserting the momentum dependent diffusons into Eq.(\ref{eqt3}). This gives
\begin{equation} \langle \langle T_a^3 \rangle \rangle=\frac{\rho_0^4}{16
\pi^2 g^2} \int {\rm d}^2{\scriptstyle Q}_1 {\rm d}^2{\scriptstyle Q}_3
\; \exp\left\{-\rho_0^2[{\scriptstyle Q}_1^2+{\scriptstyle
Q}_3^2+({\scriptstyle Q}_1+{\scriptstyle Q}_3)^2]/8 \right\} \;
F_3(|{\scriptstyle Q}_1|L,|{\scriptstyle Q}_3|L,|{\scriptstyle
Q}_1+{\scriptstyle Q}_3|L) \label{k3} .\end{equation} With
\begin{eqnarray}
F_3(x_1,x_3,x_5) &=&\sum_{\rm per} [ \frac{(x_1+x_3)^2 x_5
\cosh(x_1+x_3)
}{(x_1+x_3+x_5)^2(x_1+x_3-x_5)^2} - \frac{(x_1-x_3)^2 x_5 \cosh(x_1-x_3)
}{(x_1-x_3+x_5)^2(x_1-x_3-x_5)^2} \nonumber \\ &&- \frac{(x_1+x_3) x_5
\sinh(x_1+x_3) }{(x_1+x_3+x_5)(x_1+x_3-x_5)} + \frac{(x_1-x_3) x_5
\sinh(x_1-x_3) }{(x_1-x_3+x_5)(x_1-x_3-x_5)} \nonumber \\
&&+\frac{(x_1+x_3) \cosh(x_1+x_3+2x_5) }{4(x_1+x_3+x_5)^2}
-\frac{(x_1+x_3) \cosh(x_1+x_3-2x_5) }{4(x_1+x_3-x_5)^2} \nonumber \\
&&-\frac{(x_1-x_3) \cosh(x_1-x_3+2x_5) }{4(x_1-x_3+x_5)^2}
+\frac{(x_1-x_3) \cosh(x_1-x_3-2x_5) }{4(x_1-x_3-x_5)^2} ]\times
\nonumber \\ &&\left[x_5 \sinh(x_1) \sinh(x_3)
\sinh^2(x_5)\right]^{-1}\label{eqdefF} ,\end{eqnarray}
which is the main result in this paper.
We study again the behavior if the beam diameters are wide. In the limit of
large beam
diameter ($\rho_0 \gg  L$) one finds $F_3(0,0,0)=\frac{16}{15}$, this means
for the third cumulant $\langle \langle T_a^3 \rangle \rangle= 4
F_3(0,0,0)/3g^2$, or \begin{equation} \langle \langle T_a^3 \rangle \rangle =
\frac{16}{5} \; \langle \langle T_a^2 \rangle \rangle^2 \label{eqt3g},
\qquad \mbox{Gaussian profile;} \; \rho_0\gg L ,\end{equation} which differs
by a factor $\frac{4}{3}$ from the plane wave limit Eq.(\ref{eqt3pw}). This
is purely a geometrical effect, depending on the profile of the incoming
beam. In a real space picture this effect is best understood. The correlation
depends on the distance: it is strongest if the incoming intensities are
close. Therefore it is not surprising to see the influence of the overlap. In
Ref.\cite{probprl} this geometrical factor has been calculated for higher
orders also (the area of a Gaussian beam is defined differently there). For
the experimental relevant case that  the beam diameter is roughly equal to
the thickness, we calculated Eqs.(\ref{k3}) and (\ref{eqdefF}) numerically.
It then turns out that the behavior of Eq.(\ref{eqt3g}) is actually seen for
a large range of beam diameters. The increase of the correlation for smaller
beams, turns out to be roughly the same for both the third cumulant and the
second cumulant squared. All corrections to (\ref{eqt3g}) turn out to be
relatively small, as we will discuss below. Apart from this advantage, errors
in the sample thickness and the mean free path cancel by presenting the
results as the ratio between the second cumulant squared and the third
cumulant.

\section{Influence of internal reflection}\label{secz0}
In this section we calculate the influence of internal reflection on our
results, and show that it is small. It was seen in previous
work\cite{lisyansky,skin} that surface reflection decreases the $C_2$
correlation. In Eq.(\ref{eqt3g}) corrections from boundary reflection cancel
partly. We did not calculate the influence of internal reflections for the
general case, but only for the case of very broad beams (i.e. only for
${\scriptstyle Q}$ independent diffusons). One expects that this behavior may
be  extrapolated to the ${\scriptstyle Q}-$dependent case. At least, for the
second cumulant this is a good approximation\cite{skin}. The ${\scriptstyle
Q}-$ independent  diffusons in the presence of internal reflections
are\cite{thmn5} \begin{equation} {\cal L}^a_{\rm in} =\frac{4\pi
\tau_1(\mu_a)} {k\ell A\mu_a} \frac{L-z+z_0}{L+2z_0} , \qquad {\cal L}_{\rm
out}(z)=\frac{k}{\ell}\frac{z+z_0}{L+2z_0} ,\end{equation} here $z_0$  is the
extrapolation length. In the definition of $g$ and $T_a$ , we replace $L$
by $L+2z_0$. If internal reflections are absent $z_0$ equals $0.71 \ell$
and the corrections, which are of order $z_0/L$, are often negligible. With
internal reflection present $z_0$ increases and
should taken into account\cite{thmn5}. The correlations are known to
decrease if internal
reflections are present\cite{lisyansky,skin}. In first order of $z_0/L$ the
second  and third cumulant behave as \begin{equation} \langle \langle T_a^2
\rangle \rangle= \frac{2}{3g} \left(1-3\frac{z_0}{L}\right), \qquad \langle
\langle T_a^3 \rangle \rangle= \frac{16}{15g^2} \left(1-\frac{15}{2} \;
\frac{z_0}{L}\right) , \end{equation} Therefore, the central relation
(\ref{eqt3g}) has a correction \begin{equation} \frac{\langle \langle T_a^3
\rangle \rangle  } {\langle \langle T_a^2 \rangle \rangle^2 }= \frac{16}{5}
\left(1-\frac{3}{2} \; \frac{z_0}{L}\right)\label{eqt3z0}.  \end{equation} The
experimental determination of the index of refraction of the sample, which
determines $z_0$, is difficult\cite{denouter2}. Fortunately, the correction
is rather small for the experimental situation considered.

\section{Contributions from disconnected diagrams}\label{secjoh}
So far the leading contributions to the second and third cumulant have been
calculated. They are
given by the connected diagrams in Fig.~\ref{figt2}(b) and
Figs.~\ref{figt3}(c+d), respectively. Yet there are also contributions to the
second and third cumulant from disconnected diagrams. The diagram in
Fig.~\ref{figt2}(a) gives an additional contribution to the second cumulant,
and likewise, the diagrams of Figs.~\ref{figt3}(a) and (b) give a
contribution to the third cumulant. These disconnected diagrams correspond to
cumulant contributions which are not (fully) due to interference. They
describe effects which have little to do with the interference
effects we are after. Here we calculate their contribution and show that
they are small.

\subsection*{Extra contributions to second cumulant}
We first explain the contribution of Fig.~\ref{figt2}(a) to the second
cumulant.
As a start we use the model of a waveguide.
We assume that the disorder couples one incoming mode $a$
to all outgoing modes. A waveguide has discrete modes, and
for the moment we assume that different outgoing modes are uncorrelated. The
second
moment $ \langle T_a^2 \rangle $ is split into a connected part $\langle
T_a^2 \rangle_{\rm con}$, Fig.~\ref{figt2}(b) and a disconnected part
$\langle T_a^2 \rangle_{\rm dis}$, Fig.~\ref{figt2}(a). The total
transmission is the summation over all outgoing modes $ T_a=\sum_b
T_{ab}$.
The disconnected part of the second moment is \begin{eqnarray} \langle
T_a^2 \rangle_{\rm
dis} & =&  \sum_{b_1\neq b_2}^{N,N} \langle T_{ab_1} T_{ab_2} \rangle +
\sum_{b_1=b_2}^N \langle T_{ab_1} T_{ab_2} \rangle \nonumber \\&=& N(N-1)
\langle T_{ab}
\rangle^2 + N \langle T_{ab}^2 \rangle \nonumber \\ &=& \langle T_a \rangle^2
+ N \langle T_{ab} \rangle^2, \label{eqdis2} \end{eqnarray} where $N$ is the
number of modes supported by the waveguide. For the last equality sign the
averaged second
moment of the intensity speckle is given by the speckle distribution function
Eq.~(\ref{eqray}), $\langle T_{ab}^2 \rangle = 2 \langle
T_{ab}\rangle^2$. From Eq.(\ref{eqdis2}) we see that the disconnected
diagram $\langle T_a^2
\rangle_{\rm dis}$ does not completely factorize into the average squared
$\langle T_a \rangle^2$. Therefore it contributes to the  second cumulant.
As shown above, the connected part of the second
moment of the total transmission $\langle T_a^2 \rangle_{\rm con}$ is
proportional to $L/N\ell$. For the sum of the disconnected and the connected
contribution to the second cumulant one thus finds
\begin{equation} \langle
\langle T_a^2 \rangle\rangle =\langle \langle T_a^2 \rangle\rangle_{\rm con}
+\langle \langle T_a^2 \rangle\rangle_{\rm dis} = \frac{L}{2N\ell} +
\frac{1}{N}, \end{equation}
which also holds for plane wave case. After this we turn to the situation of
a diffusely scattering slab, with a finite focus of the incoming beam.  The
incoming beam will be broadened in the transverse direction by diffusion,
changing the above result. To calculate the different contributions to the
cumulants, for the moment the intensity distribution at the exit interface at
transversal coordinates ${\scriptstyle R}_1$ and ${\scriptstyle R}_2$ is
needed.

The amplitudes, making up each diffuson, can propagate from to outgoing
surface in different directions, respectively ${\scriptstyle P}_1$,
${\scriptstyle P}_2$, ${\scriptstyle P}_3$ and ${\scriptstyle P}_4$.
\begin{equation}
\Psi({\scriptstyle R}_1){\rm e}^{i{\scriptstyle
P}_1{\scriptstyle R}_1}
\Psi^*({\scriptstyle R}_1) {\rm e}^{-i{\scriptstyle P}_2
{\scriptstyle R}_1} \Psi({\scriptstyle R}_2){\rm
e}^{i{\scriptstyle P}_3{\scriptstyle R}_2}
\Psi^*({\scriptstyle R}_2){\rm e}^{-i{\scriptstyle P}_4
{\scriptstyle R}_2}. \end{equation}
To get the contribution to the second moment of the total
transmission, one first integrates over the transversal
coordinates ${\scriptstyle R}_1$ and ${\scriptstyle R}_2$ to get the
contribution of the whole exit interface to the intensity in a
certain direction, then one integrates over all directions to get the
total transmission. To obtain intensities the amplitudes need to be paired
giving the
following possibilities. The first possible pairing of the amplitudes is
${\scriptstyle P}_1={\scriptstyle P}_2$, ${\scriptstyle P}_3={\scriptstyle
P}_4$, see Fig.~\ref{johfig1}(a), and brings \begin{equation}
 \int {\rm d}{\scriptstyle R}_1 \;
 \int {\rm d}{\scriptstyle R}_2 \; \langle
\Psi({\scriptstyle R}_1) \Psi^*({\scriptstyle R}_1)
 \Psi({\scriptstyle
R}_2) \Psi^*({\scriptstyle R}_2) \rangle = I^2(0),
\end{equation}
where we have defined $I({\scriptstyle Q})$ as the transmission by a diffuson
with transverse momentum ${\scriptstyle Q}$. Including
the incoming Gaussian beam profile, it is proportional to, see
Eq.(\ref{eqladQ}),(\ref{eqpsig}) \begin{equation}
 I({\scriptstyle Q}) \propto \frac{  |{\scriptstyle Q}|  {\rm
e}^{-{\scriptstyle Q}^2
\rho_0^2/8}}{ \sinh|{\scriptstyle Q}| L} \label{iqspace}.
\end{equation}
Integrating over all outgoing directions results in
\begin{equation}
\frac{\langle T_a^2 \rangle}{\langle T_a \rangle^2} = \frac{\pi^2 k^4
I^2(0)}{ \pi^2 k^4 I^2(0)  }.
\end{equation}
This is the just the factorizing contribution, and hence does not
contribute to the second cumulant.

The second possible pairing of the amplitudes, ${\scriptstyle
P}_1={\scriptstyle
P}_4$, ${\scriptstyle P}_2={\scriptstyle P}_3$, does give a
contribution to the second
cumulant. It is the diagram in  Fig.~\ref{johfig1}(b). \begin{eqnarray}
 \int {\rm d}{\scriptstyle R}_1
{\rm d}{\scriptstyle R}_2 \;
\langle \Psi({\scriptstyle R}_1) \Psi^*({\scriptstyle
R}_1) \Psi({\scriptstyle R}_2)
\Psi^*({\scriptstyle R}_2) \rangle
{\rm e}^{i{\scriptstyle R}_1({\scriptstyle P}_1-{\scriptstyle P}_2)} {\rm
e}^{-i{\scriptstyle R}_2({\scriptstyle P}_1-{\scriptstyle P}_2)}  &=&
I({\scriptstyle P}_1-{\scriptstyle P}_2) I({\scriptstyle P}_2-{\scriptstyle
P}_1). \end{eqnarray}
Subsequent integration over the outgoing directions $ {\scriptstyle P}_1 $
and $ {\scriptstyle P}_2 $ yields
\begin{equation}
\langle T_a^2 \rangle = \pi k^2 \int_{|{\scriptstyle Q}|<k} {\rm d}^2
{\scriptstyle Q} \; I^2({\scriptstyle Q}). \end{equation}
The integral can be extended to infinity since $I({\scriptstyle Q})$ is an
exponentially decaying function. The contribution of the disconnected diagram
Fig.~\ref{figt2}(b) to the second cumulant is thus \begin{equation} \langle
\langle T_a^2 \rangle \rangle_{\rm dis} = \frac{\pi k^2 \int_{-\infty}^\infty
{\rm d}^2{\scriptstyle Q} I^2({\scriptstyle Q}) } {\pi^2 k^4 I^2(0)}\equiv
\frac{1}{N} \label{modesdef} \end{equation} In analogy to the waveguide, this
result describes irreducible contributions from disconnected diagrams. It can
be interpreted as the inverse of the number of independent speckle spots in
transmission at the exit interface \cite{goodmanboek,t3prl}.

\subsection*{Extra contributions to the third cumulant}

We apply the same method for contributions to the third cumulant. Following
the  waveguide argument as above, we find the contributions to the third
cumulant of the diagrams of Figs.~\ref{figt3}(c+d), \ref{figt3}(b), and
\ref{figt3}(a), respectively, \begin{equation} \langle \langle T_a^3\rangle
\rangle=\frac{3}{5} \frac{L^2}{N^2 \ell^2} + \frac{6}{2} \frac{L}{N^2 \ell} +
\frac{2}{N^2} \label{cum3reducapprox}. \end{equation} The first r.h.s. term
is the connected diagram. It is clear that the second term, the diagram
Fig.~\ref{figt3}(b), gives a much larger contribution to the third cumulant
than the third term, the diagram of Fig.~\ref{figt3}(a), as the diagram
Fig.~\ref{figt3}(b) is
already enhanced by some interference. In the following we only consider this
diagram.

As can be seen from Fig.~\ref{figt3}(b) there are three possibilities
to combine the three diffusons into two connected diffusons.
and a single diffuson. Attaching outgoing directions
to the amplitudes at the exit interface gives (see Fig.~(\ref{johfig2}),
\begin{equation}  \Psi({\scriptstyle R}_1){\rm
e}^{i{\scriptstyle P}_1{\scriptstyle R}_1} \Psi^*({\scriptstyle
R}_1) {\rm e}^{-i{\scriptstyle P}_2
{\scriptstyle R}_1} \Psi({\scriptstyle R}_2){\rm
e}^{i{\scriptstyle P}_3{\scriptstyle R}_2} \Psi^*({\scriptstyle
R}_2){\rm e}^{-i{\scriptstyle P}_4
{\scriptstyle R}_2} \Psi({\scriptstyle R}_3) {\rm
e}^{i{\scriptstyle P}_5 {\scriptstyle R}_3}
\Psi^*({\scriptstyle R}_3) {\rm e}^{-i{\scriptstyle P}_6 {\scriptstyle R}_3}.
\end{equation}
It is clear that there are six possibilities to pair the outgoing
directions into intensities.
Integrating over the transversal coordinates ${\scriptstyle R}_1,
{\scriptstyle R}_2$ and ${\scriptstyle R}_3$
gives the following six contributions,
\begin{eqnarray}& I_{\rm con}(0)I_{\rm con}( 0) I( 0) & \quad I_{\rm
con}({\scriptstyle P}_1-{\scriptstyle P}_3) I_{\rm con}({\scriptstyle
P}_3-{\scriptstyle P}_1) I( 0) \nonumber \\ & \qquad I_{\rm
con}( 0) I_{\rm con}({\scriptstyle P }_1-{\scriptstyle P}_5) I({\scriptstyle
P}_5-{\scriptstyle P}_1) &\quad  I_{\rm con}(
0) I_{\rm con}({\scriptstyle P}_3-{\scriptstyle P}_5) I({\scriptstyle
P}_5-{\scriptstyle P}_3) \nonumber \\ &\qquad \qquad I_{\rm
con}({\scriptstyle P}_1-{\scriptstyle P}_5) I_{\rm con}({\scriptstyle
P}_3-{\scriptstyle P}_1) I({\scriptstyle P}_5-{\scriptstyle P}_3)& \quad
I_{\rm con}({\scriptstyle P}_1-{\scriptstyle P}_3) I_{\rm
con}({\scriptstyle P}_3-{\scriptstyle P}_5) I({\scriptstyle
P}_5-{\scriptstyle P}_1 ) \label{eqt3pair}  ,\end{eqnarray}
where $I_{\rm con}$ denotes the transmitted intensity coming from the
connected part of diagram Fig.~\ref{figt3}(b).
The third cumulant is in the discrete mode model given by
\begin{equation}
\langle \langle T_a^3 \rangle \rangle =
 \frac{1}{\langle T_{a} \rangle^3}\left[
\sum_{b,b',b''} \langle T_{ab}T_{ab'}T_{ab''} \rangle - 3
\sum_b \langle T_{ab} \rangle
\sum_{b,b'}\langle T_{ab}T_{ab'} \rangle + 2 \sum_b \langle T_{ab}
\rangle^3 \right] \end{equation}
Similarly to the second cumulant we insert all pairings
of Eq.~(\ref{eqt3pair})
and find for the contribution from disconnected diagrams \begin{equation}
\langle \langle T_a^3 \rangle\rangle_{\rm dis}= \frac{6}{\langle
T_a\rangle^3} \int {\rm d}^2{\scriptstyle Q}_1 d^2{\scriptstyle Q}_2
d^2{\scriptstyle Q}_3
\left[ I_{\rm con}( 0) I_{\rm con}({\scriptstyle Q}_1) I(-{\scriptstyle
Q}_1) + I_{\rm con}({\scriptstyle Q}_1) I_{\rm con}( {\scriptstyle Q}_2)
I({\scriptstyle Q}_3) \right].
\end{equation}
The integrand is dominated by its first term, which is depicted in
Fig.~\ref{johfig2}. We explicitly evaluated
this term. The calculation follows completely the line of the second cumulant
calculation, yet it is slightly more complicated as it has a
${\scriptstyle Q}-$dependent
outgoing diffuson. The integration over ${\scriptstyle Q}_2$ and
${\scriptstyle Q}_3$ gives a factor $(\pi k^2)^2$. Comparing the result
with the last two terms in third cumulant
in the waveguide model, Eq.~(\ref{cum3reducapprox}), we define
$N^*$ as \begin{equation} \langle \langle T_a^3 \rangle \rangle_{\rm dis} =
 6 \langle \langle T_a^2 \rangle \rangle_{\rm con} / N^* .
\end{equation}
The number $N^*$ is inversely proportional to the contribution of
disconnected
diagrams to the third cumulant for a Gaussian profile.

\subsection*{Polarization effects}
The vector character of the light has not been taken into account yet. The
two independent polarizations of each outgoing direction effectively double
the number of independent speckle spots $N$. Thus for an incoming plane wave
of unit intensity with fixed polarization the total transmission and the
conductance $g$ are twice as large as they would be in the scalar case. As
we work with normalized cumulants, this effect reduces only the value of
the second cumulant ($\propto 1/g$) and the value of the third cumulant
($\propto 1/g^2$). Therefore, it is immediately seen that the relation
$\langle \langle T_a^3 \rangle \rangle \propto \langle \langle
T_a^2 \rangle \rangle ^2 $ (Eq.~\ref{eqt3g}) is not affected.
The vector character does reduce the correction Eq.~(\ref{modesdef}) by a
factor ${\it 2}$.
For the experimental data of Ref.~\cite{t3prl} the number of modes $N$ as
well as $N^*$ is listed in Table~\ref{sampletab} (including the doubling).

Summarizing the previous sections we have included three
corrections. We first obtained the result for very broad Gaussian beams, in
the large $L/\ell$ limit, Eq.~(\ref{eqt3g}). The first correction was
the influence of a finite beam diameter, it changes the diffuse intensity
from linear into an exponentially decaying, see Eq.~(\ref{eqladQ}).
This correction is contained in  Eq.~(\ref{eqdefF}) and Eq.~(\ref{k3}).
The presence of internal reflections also changes the spatial dependence of
the diffuse intensity resulting in a correction Eq.~(\ref{eqt3z0}).
The third correction is of another nature, it is the only process which does
not come from interference, but from disconnected diagrams. Only this
term depends the number of modes, which in the vector case is twice as
large as in the scalar case.

\section{Comparison with experiments}\label{secres}

The data set found experimentally in Ref.\cite{t3prl} is reproduced in Table
\ref{sampletab} and Fig.~\ref{figresult}. The experiments reported there were
performed with seven different samples. The experimental setup and
measurement technique used is extensively described in Ref.\cite{deboer}.
Samples consisted of 36 vol.\% rutile $ TiO_{2}$ pigment on a transparent
substrate. The extrapolation length was estimated from the effective index of
refraction to be $z_0\approx 1.1 \mu m$. The absorption length $\ell_{a}$ was
determined to be $\simeq 70 \mu m$. Different values of the conductance $g$
were probed by taking various sample thicknesses and by varying the beam
diameter. The fluctuations in the total transmission were measured by
varying the wavelength of the light.

For a very broad beam we found the simple relation (\ref{eqt3g}) between
second and third cumulant. A weighted least square fit to \begin{equation}
\langle \langle T_a^3 \rangle \rangle=const. \langle \langle T_a^2 \rangle
\rangle^2 \end{equation} of the raw experimental data yields a prefactor
$2.9\pm
0.6$.
However, as discussed above, there are three corrections to be made.
First, if the beam width becomes comparable to sample thickness the integrals
(\ref{c2}) and (\ref{k3}) have to be performed: If the beam width reduces,
$g$ decreases accordingly and both cumulants increase in absolute size. Yet
the precise increase is somewhat different resulting in a somewhat smaller
prefactor in Eq. (\ref{eqt3g}). We corrected each data point individually for
its finite focus, mapping it to the infinite focus case. The third cumulant
was multiplied by a factor which ranged from 1.03 to 1.13, as $L/\rho_0$
ranged from 0.41 to 6.3, see Table~\ref{sampletab}. This is the largest
correction, it changes the prefactor some 10\%. Secondly, we corrected for
internal reflections according to Eq.(\ref{eqt3z0}). The third correction
comes from the disconnect diagrams. The contributions from the
disconnected diagrams are substracted from the measured
cumulants. After all these corrections the
data should again obey the law: $\langle \langle T_a^3 \rangle \rangle=3.2 \,
\langle \langle T_a^2 \rangle \rangle^2$. The results are plotted in Fig.\
\ref{figresult}, where the points are the corrected data points and the line
is the theoretical prediction. A least square fit gives \begin{equation}
\langle \langle T_a^3 \rangle \rangle=(3.3 \pm 0.6) \langle \langle T_a^2
\rangle \rangle^2. \end{equation} Note that there is no adjustable parameter.
We find that there is good agreement between experiment and theory. All
corrections are minor as compared to the error in the data, fitting the raw
data is also in agreement with the theoretical value of 3.2. We recall that
the major shift between fits of raw and corrected data comes from the beam
diameter coreection. Inspecting the figure one might be tempted to make a
linear fit, but in Ref.\cite{t3prl} it was shown that this fit is
statistically very improbable.

\section{Discussion}
We have calculated the second and third cumulant of the distribution of the
total transmission underlying the conclusions of Ref.~\cite{t3prl} and
compared it to the experimental data. Both cumulants are a consequence of
interference between diffuse channels. They were calculated with a
diagrammatic technique. The inverse dimensionless conductance, interpreted as
an interference probability, is a perturbation parameter in the theory. The
third cumulant is proportional to the second cumulant squared. We also found
a non-trivial dependence on the profile of the incoming beam used. The
cumulants were calculated for arbitrary beam diameter, but the  influence of
a finite focus on the ratio is rather weak. Also boundary reflections were
included. Our calculations confirm that the main contributions come from
diagrams with interference processes, i.e. connected diagrams, as we have
shown that the contributions from disconnected diagrams is small. The
experimentally found ratio of the third cumulant versus the second cumulant
squared is well described by our theory.

The extension of the calculations to higher cumulants is straightforward. The
$n$-th cumulant will contain $(n-1)$ Hikami four-point vertices.  So the
contribution is estimated to be: $\langle \langle T_a^n \rangle \rangle
\propto g^{1-n}$. Also corrections and cancellations from higher order
vertices are present, but it is clear that the calculation becomes very
laborious at large $n$. Recently two of the authors discovered that all the
cumulants of the distribution function can be mapped onto the moments of the
eigenvalue distribution of the transmission matrix\cite{probprl}. The
eigenvalue distribution is bimodal and was first calculated using random
matrix techniques\cite{dorokhov} but recently its validity beyond quasi-1D
was proven\cite{nazarov}. As the eigenvalue distribution is known, the entire
distribution of the total transmission was calculated in the limit of broad
beams. These results agree with calculations presented here for the first
three cumulants. The experimental data thus also proof the first
few moments of the eigenvalue distribution function. As only three moments
are known, it is impossible to reconstruct the full eigenvalue distribution
from the experimental data. The present calculation leads us to assume that
the eigenvalue
distribution in a diagrammatic approach is also given by loopless connected
diagrams. As the ratio of first few cumulants does not depend sensitively on
the beam diameter, the results of Ref.\cite{probprl} are probably also valid
in the regime where the beam diameter becomes comparable to the sample
thickness.

\subsection{Acknowledgements}
The authors thank E. Kogan and C. W. J. Beenakker for discussion. This work
was partly supported by the Stichting vor Fundamenteel Onderzoek der Materie
(FOM), which is a part of the Nederlandse Organisatie voor Wetenschappelijk
Onderzoek. The research of Th.M. Nieuwenhuizen was supported by the Royal
Netherlands Academy of Arts and Sciences (KNAW) and was also sponsored by
NATO (grant nr. CGR 921399).

\newpage

\begin{table}
\caption{Sample thickness, beam width,
second cumulant, third cumulant for
the different samples as taken from
Ref.\protect{\cite{t3prl}}. Next are the beam diameter correction factor on
the third cumulant (see section \protect{\ref{secres}}), and the number of
modes $N$ and $N^*$.
Together they give the corrected (plotted) cumulants (last two columns).}
\label{sampletab} \begin{tabular}{rrccrrrrrr}
Sample     & Beam           & Second &Third&   Beam& Number&Number&
Corrected &Corrected \\ thickness L &diameter $\rho_0$ &cumulant& cumulant
&diameter & of modes&of modes&2e cum.& 3th cum.\\
 (in $\mu m$) & (in $\mu m$)  & $(\times 10^{-4})$ &
$(\times 10^{-7})$ &correction & $N$ & $N^*$& $(\times 10^{-4})$&$(\times
10^{-7})$
\\ \hline 30 & 77 &0.36$\pm$0.01 &0.014 $\pm$ 0.035& 1.03& 388000& 300000&
0.33& 0.007\\
12 & 26  &0.97 $\pm$0.03&$-$0.03 $\pm$0.25&1.04 & 46800
&36800& 0.75& $-$0.18 \\
22 & 32  &1.24 $\pm$0.04&0.68 $\pm$0.28   &1.06  &88600 &72100
&1.13&0.58\\
30 & 33  &1.57 $\pm$0.04&1.30 $\pm$0.46   &1.07 &119000 &99300
&1.49 &1.21\\
53 & 35  &1.80 $\pm$0.03&0.91 $\pm$0.53   &1.10  &241000 &212000
&1.76 &0.86  \\
30 & 26  &1.90 $\pm$0.03&0.92 $\pm$0.56   &1.09 &94900 &81300
&1.79 &0.78  \\
45 & 33  &1.90 $\pm$0.05&1.33 $\pm$0.43   &1.10 &187000 &163000
&1.84&1.26   \\
53 & 26  &2.18 $\pm$0.03&1.77 $\pm$0.59   &1.10 &208000 &189000
&2.13&1.70 \\
170& 27  &2.69 $\pm$0.06&2.02 $\pm$0.82   &1.13 &1420000&1430000
&2.68 &2.00 \\
78 & 28  &2.74 $\pm$0.03&2.43 $\pm$0.62   &1.11 &396000&372000
&2.71 & 2.39  \\
30 & 17  &4.82 $\pm$0.10&9.1 $\pm$3.3     &1.11 &71900&64400
&4.68&8.60   \\
30 & 10  &8.01 $\pm$0.36&5.3 $\pm$6.4     &1.11 &60200&56400
&7.84&4.47
\end{tabular}
\end{table}

\newpage

\begin{figure}
\caption{Left: an example of an actual scattering process; a retarded (full
line) and an advanced amplitude (dashed line) come from the left and share
the same path through the sample. Right: schematic representation of the
average process, the diffuson (for clarity the scatterings are not drawn).}
\label{figdif} \end{figure}

\begin{figure}
\caption{The two contributions to the second moment of the total
transmission. In diagram (a) the transmission channels are independent; this
process is of order unity and is almost completely reducible to the mean
value squared. Diagram (b) corresponds to two interfering channels; this is
the second cumulant, it is of order $1/g$. The close parallel lines are
diffusons; the shaded square denotes the Hikami four point
vertex.}\label{figt2} \end{figure}

\begin{figure}[h]
\caption{The three contributions to the third moment of the total
transmission. Diagram a) corresponds to independent transmission channels; it
is of order 1. In diagram b) there is correlation but it can be decomposed
into the second cumulant; this is of order $g^{-1}$. The diagrams c) and d)
are the contributions to the third cumulant, $O(g^{-2})$. }\label{figt3}
\end{figure}

\begin{figure}
\caption{The Hikami four point vertex. It describes the exchange of amplitudes
of two incoming diffusons $a$ and $a'$ into two outgoing diffusons $b$ and
$b'$. The dots linked with the dashed line denote the dressing with an extra
scatterer. }\label{figh4sob} \end{figure}

\begin{figure}[h]
\caption{Diagrams contributing to the interaction of six diffusons: $H_6$.
To 1,3,5 the incoming diffusons are connected, to 2,4,6 the outgoing ones.
Possible rotations of the three rightmost diagrams are not drawn, in
total there are sixteen diagrams.} \label{figh6}
\end{figure}

\begin{figure}[t]
\caption{The disconnected contribution to the second moment.
In (a) the two transmissions factorize into the average value squared, and
does not contribute to the second cumulant. Diagram (b) is much smaller, but
gives a contribution to the second cumulant.
The amplitudes making up the two diffusons propagate out in different
directions.}\label{johfig1}
\end{figure}

\begin{figure}[t]
\caption{The leading disconnected contribution to the third
cumulant, the box symbolises again the Hikami vertex.}
\label{johfig2}
\end{figure}

\begin{figure}
\caption{The third cumulant plotted against the second cumulant. The  points
are the corrected experimental data (see text). The line is the theoretical
prediction, $\langle \langle T_a^3 \rangle \rangle=3.2 \langle \langle T_a^2
\rangle \rangle $; no free parameters were introduced.} \label{figresult}
\end{figure}

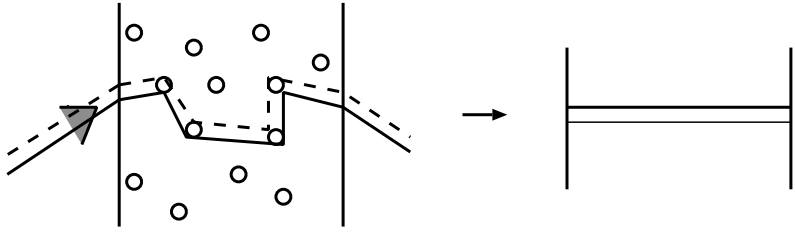
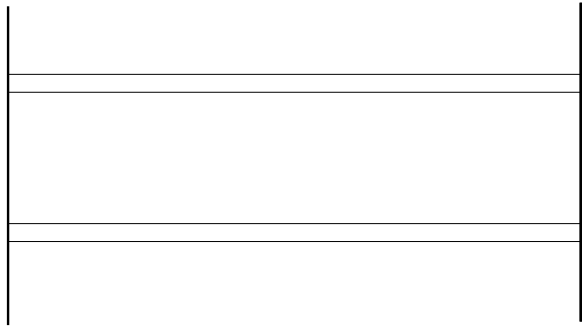
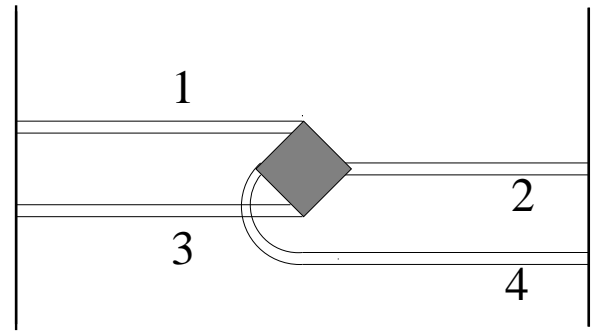
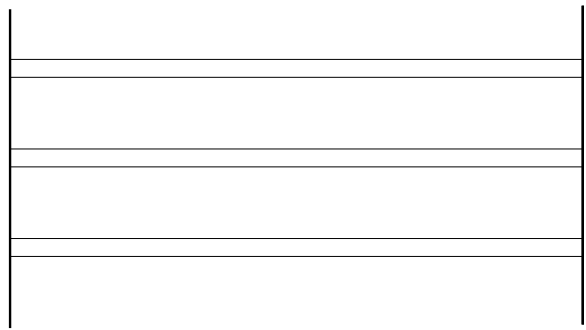
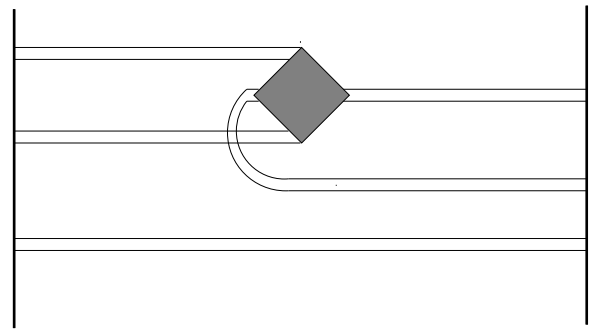
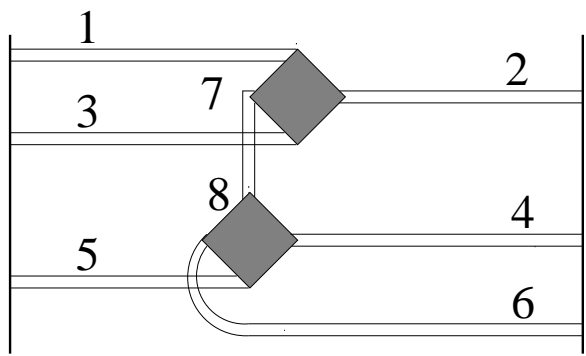
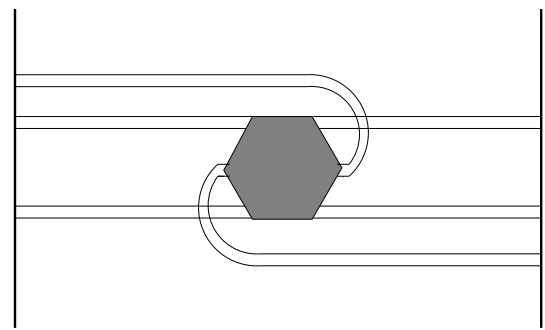
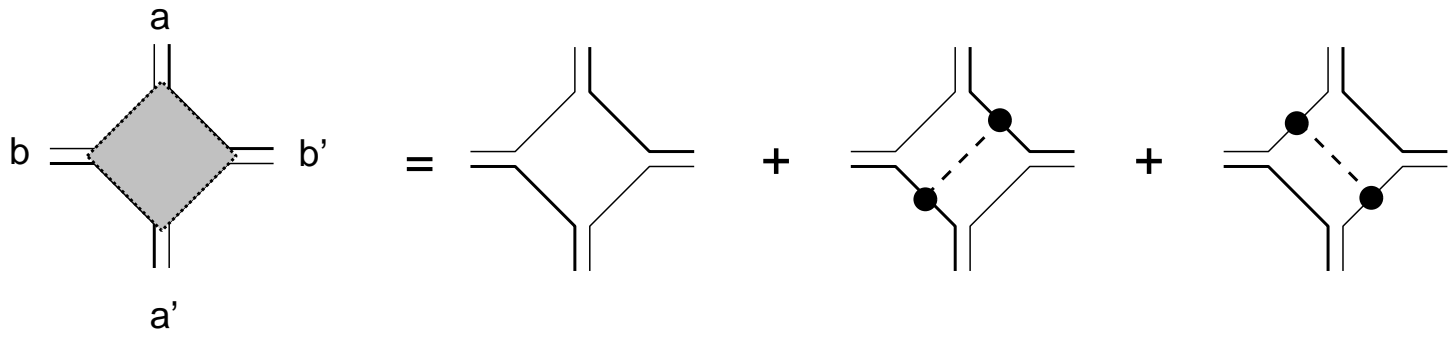
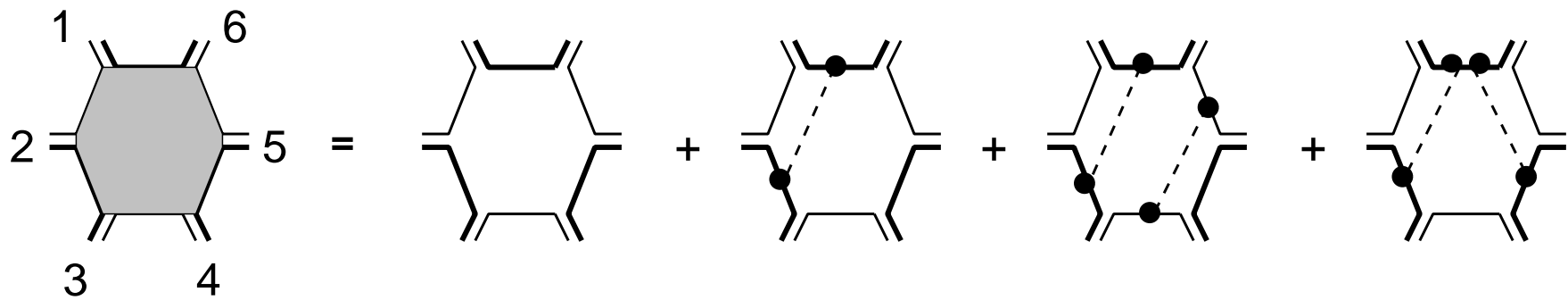
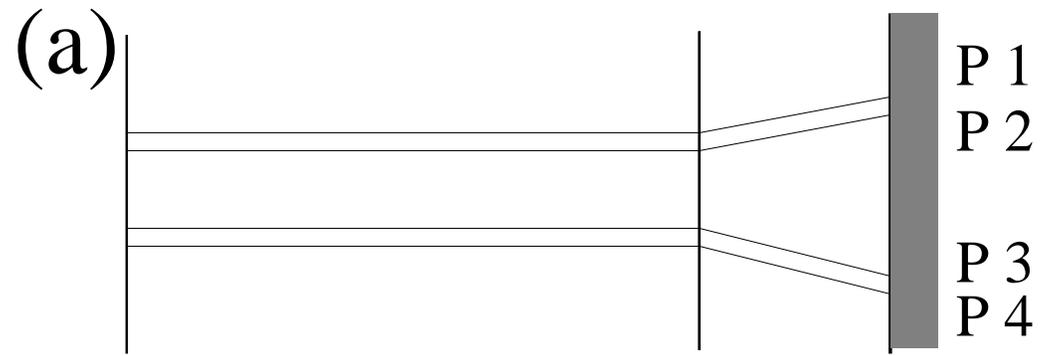
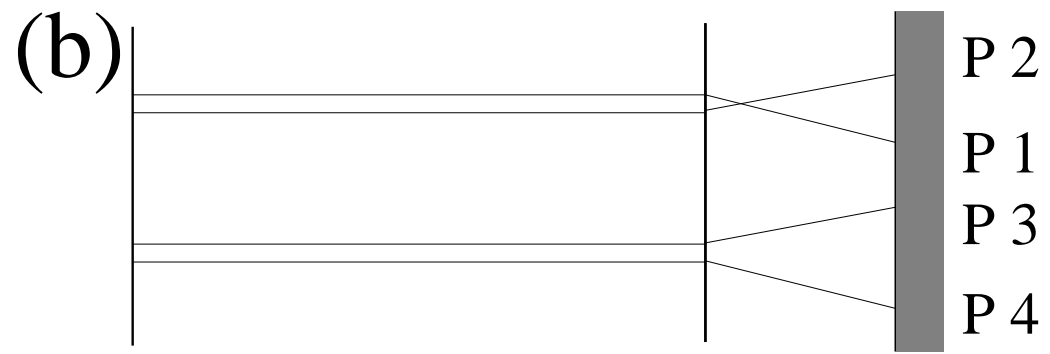
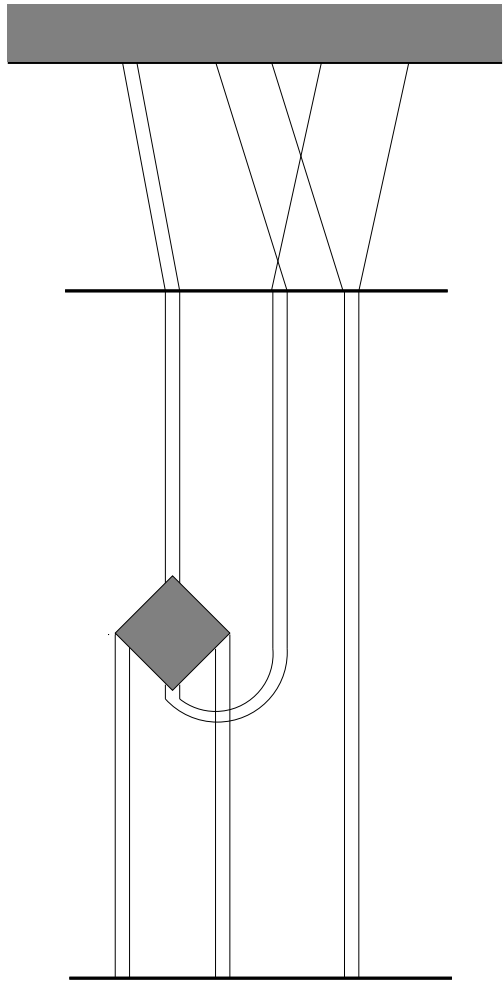
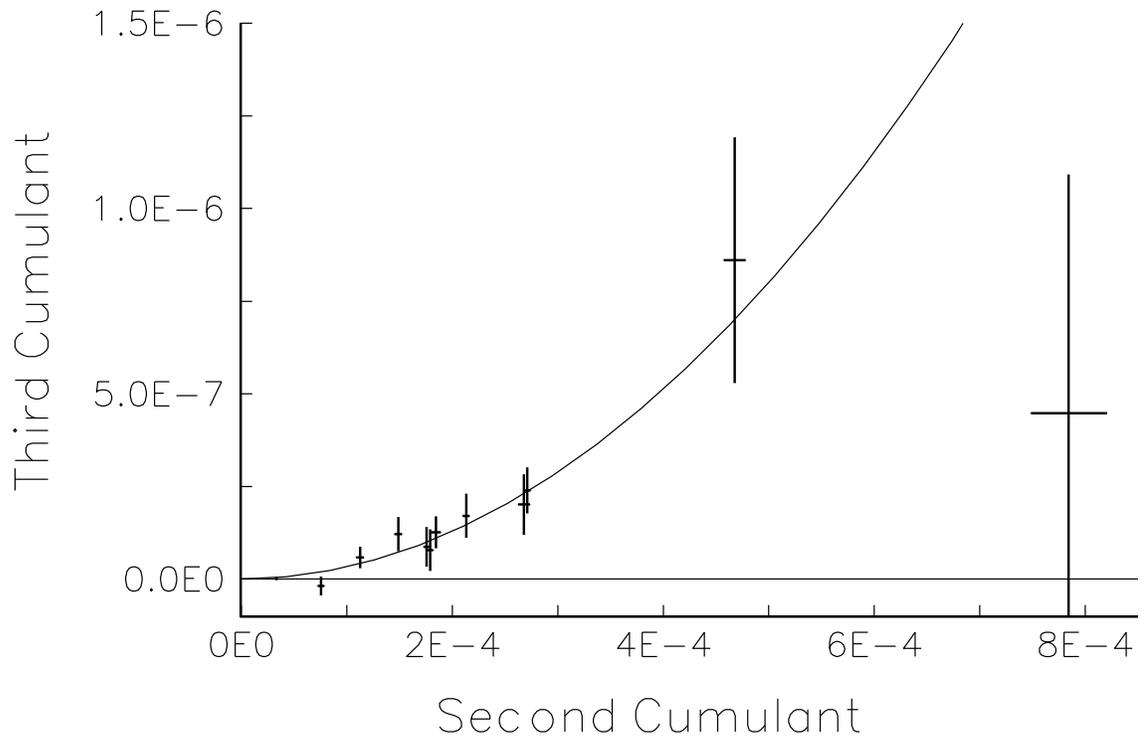
\end{document}